\documentclass[english,twocolumn,prd,superscriptaddress,showpacs,nofootinbib,preprintnumbers]{revtex4}
\usepackage[T1]{fontenc}
\usepackage[latin1]{inputenc}
\usepackage{graphicx}

\makeatletter

\makeatletter

\makeatletter

\makeatletter

\makeatletter

\makeatletter

\usepackage{amsfonts}

\usepackage{dcolumn}

\def\be{\begin{equation}}
\def\ee{\end{equation}}
\def\ba{\begin{eqnarray}}
\def\ea{\end{eqnarray}}
\def\bs{\begin{subequations}}
\def\es{\end{subequations}}

\usepackage{color}

\def\textrm{\rm}

\makeatother

\makeatother

\makeatother

\makeatother

\makeatother

\usepackage{babel}
\makeatother
\begin{document}

\title{Are $f(R)$ dark energy models cosmologically viable ?}

\author{Luca Amendola}

\affiliation{INAF/Oss. Astronomico di Roma, Via Frascati 33, 00040 Monte Porzio
Catone (Roma), Italy}

\author{David Polarski}

\affiliation{LPTA, Universit\'{e} Montpellier II, UMR 5207, 34095 Montpellier
Cedex 05, France}

\author{Shinji Tsujikawa}

\affiliation{Department of Physics, Gunma National College of Technology, Gunma
371-8530, Japan}

\date{\today{}}

\begin{abstract}
All $f(R)$ modified gravity theories are conformally identical to
models of quintessence in which matter is coupled to dark energy with
a strong coupling. This coupling induces a cosmological evolution
radically different from standard cosmology. We find that in all $f(R)$
theories that behave as a power of $R$ at large or small $R$ (which
include most of those proposed so far in the literature) the scale
factor during the matter phase grows as $t^{1/2}$ instead of the
standard law $t^{2/3}$. This behaviour is grossly inconsistent with
cosmological observations (e.g. WMAP), thereby ruling out these models
even if they pass the supernovae test and can escape the local gravity
constraints. 
\end{abstract}


\maketitle
The late time accelerated cosmic expansion is a major challenge to
cosmology \cite{review}. It can be due to an exotic component with
sufficiently negative pressure, Dark Energy (DE), or alternatively
to a modification of gravity, no longer described by General Relativity.
Examples of such modified gravity DE models are theories where the
Ricci scalar $R$ in the Lagrangian is replaced by some function $f(R)$,
e.g. inverse powers $R^{-n}$ \cite{Capo,Carroll}. Although these
models exhibit a natural acceleration mechanism, criticisms emphasized
their inability to pass solar system constraints \cite{local}. Indeed,
$f(R)$ theories correspond to scalar-tensor gravity with vanishing
Brans-Dicke parameter $\omega_{\textrm{BD}}$ \cite{TT83}. However
one could in principle build models with a very short interaction
range (e.g. adding a $R^{2}$ term \cite{ON,brook}) or assume decoupling
of the baryons from modified gravity. Since these models could pass
local gravity constraints, it is important to assess their cosmological
viability: this is the aim of this Letter. We will consider models
of the form $f(R)=R-\mu^{2(n+1)}/R^{n}$ for all $\mu,n$ such that
$df/dR>0$; for all these models, the scale factor $a(t)$ expands
as $t^{1/2}$ instead of the conventional $t^{2/3}$ behaviour during
the matter phase that \emph{precedes} the final accelerated stage
(in contrast with \emph{inflationary} models like Starobinsky's $R^{2}$
one \cite{Sta80}). This would lead to inconsistencies with the observed
distance to the cosmic microwave background (CMB), the large scale
structure (LSS) formation, and the age of the Universe. This crucial
fact appears to have been overlooked so far.

Consider the general action in the Jordan frame (JF) \begin{eqnarray}
S=\int{\textrm{d}}^{4}x\sqrt{-g}\left[\frac{1}{2\kappa^{2}}f(R)+{\mathcal{L}}_{m}\right]\,,\end{eqnarray}
 where $\kappa^{2}\equiv8\pi G$ ($G$ is the gravitational constant).
For a flat Friedmann-Robertson-Walker (FRW) metric the equations are
given by \begin{eqnarray}
3FH^{2} & = & (RF-f)/2-3H\dot{F}+\kappa^{2}~\rho_{m}\,,\nonumber \\
2F\dot{H} & = & -\ddot{F}+H\dot{F}-\kappa^{2}~(\rho_{m}+p_{m})\,,\label{Jeq1}\end{eqnarray}
 where $F\equiv\partial f/\partial R$, $H\equiv\dot{a}/a$, and $\rho_{m}$
and $p_{m}$ represent the energy density and the pressure of a perfect
fluid, obeying the standard conservation equation. These equations
coincide with a scalar-tensor Brans-Dicke theory with a potential
and vanishing $\omega_{{\rm BD}}$ \cite{BEPS00,EP00}.

Under the conformal transformation $\widetilde{g}_{\mu\nu}=e^{2\omega}g_{\mu\nu}$,
$2\omega=\log F$, one obtains the Einstein frame (EF)action: \begin{eqnarray}
S_{E}=\int{\textrm{d}}^{4}x\sqrt{-\widetilde{g}}\left[\frac{R(\widetilde{g})}{2\kappa^{2}}-\frac{1}{2}(\widetilde{\nabla}\phi)^{2}-V(\phi)+\widetilde{\mathcal{L}}_{m}(\phi)\right],\end{eqnarray}
 where $\phi\equiv\sqrt{6}\omega/\kappa$ and $V=\textnormal{sign}(F)(RF-f)/2\kappa^{2}F^{2}$
(all tilded quantities are in EF). The conformal transformation is
singular for $F=0$, so we will consider only positive-definite forms
of $F$. Quantities in the two frames are related as follows \begin{eqnarray}
\widetilde{\rho}_{m}=\rho_{m}e^{-4\omega},~~\widetilde{p}_{m}=p_{m}e^{-4\omega},~~{\textrm{d}}\widetilde{t}=e^{\omega}{\textrm{d}}t,~~\widetilde{a}=e^{\omega}a.\label{relation}\end{eqnarray}
Although we will work mainly in the EF, we checked all numerical and
analytical results directly in the JF as well. In EF the field $\phi$
and the fluid satisfy the standard gravitational and conservation
equations: \begin{eqnarray}
 &  & \ddot{\phi}+3\tilde{H}\dot{\phi}+V_{,\phi}=\sqrt{2/3}\,\kappa\beta(\widetilde{\rho}_{m}-3\widetilde{p}_{m}),\label{eq:phi}\\
 &  & \dot{\widetilde{\rho}}_{m}+3\widetilde{H}\left(\widetilde{\rho}_{m}+\widetilde{p}_{m}\right)=-\sqrt{2/3}\,\kappa\beta\dot{\phi}(\widetilde{\rho}_{m}-3\widetilde{p}_{m}),\label{eq:rhom}\end{eqnarray}
 where the coupling $\beta$ is given by \begin{eqnarray}
\beta=1/2\,,\label{beta}\end{eqnarray}
 \emph{regardless of the form} of $f(R)$. Then the strength of the
coupling between the field and the fluid is uniquely determined in
all $f(R)$ gravity theories. A dimensionless strength of order unity
means that matter feels an additional scalar force as strong as gravity
itself. Note that $\beta$ is related to $\omega_{\textrm{BD}}$ via
the relation $\beta=[3/4(2\omega_{\textrm{BD}}+3)]^{1/2}$. The dynamics
of the system depends upon the form of the potential $V(\phi)$, i.e.,
the choice of $f(R)$. For theories in which $f(R)=-\mu^{2(n+1)}R^{-n}$
($n\not=-1,0$, negative $n$ are also included), the potential in
EF is a pure exponential \begin{eqnarray}
V(\phi)=A\exp\left(-\lambda\kappa\phi\right)\,\,,\label{exppo}\end{eqnarray}
 where $\lambda=\frac{\sqrt{6}}{3}\frac{n+2}{n+1}$ and $A=\frac{\mu^{2}(n+1)}{2\kappa^{2}n|n|^{1/(n+1)}}$.
The condition $F>0$ implies $A>0$ except for $-1<n<0$: in this
case since the potential becomes negative we analyse directly the
JF. In EF the $R^{-n}$ model corresponds to a coupled DE scenario
studied in Refs.~\cite{Luca1,BNST} with the coupling (\ref{beta}).
We first discuss the main properties of this exponential potential
and then extend them to the general case.

\begin{figure}
\includegraphics[%
  bb=100bp 280bp 498bp 800bp,
  clip,
  scale=0.4]{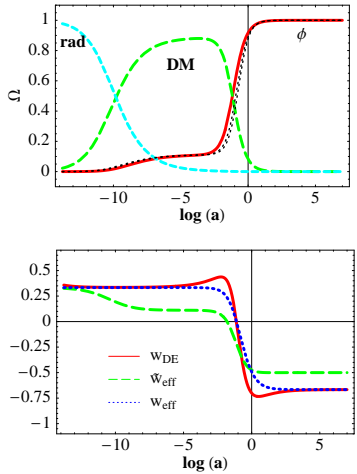}

\caption{Evolution of the fractional energy densities ($\phi$, matter, radiation)
in EF for the model $f(R)=R-\mu^{4}/R$ (top panel). Overimposed as
dotted lines the evolution of $\tilde{\Omega}_{\phi}$ for $n=4$
and $10$. Notice the constant value $\tilde{\Omega}_{\phi}\simeq1/9$
in the $\phi$MDE phase between radiation and DE domination. In the
bottom panel we plot the evolution of the observed EOS $w_{\textrm{DE}}$
of DE in JF and the effective EOS in both EF and JF ($n=1$). }

\begin{center}\label{won1}\end{center}
\end{figure}

As shown in \cite{BNST} for all values of $n$ outside $(-1,0)$
the system has one and only one global attractor solution, a scalar-field
dominated solution with an energy fraction $\widetilde{\Omega}_{\phi}=1$.
This solution appears when the potential term in Eq.~(\ref{eq:phi})
dominates over the coupling term on the r.h.s., and is therefore independent
of the coupling. On this attractor the scale factor evolves as $\tilde{a}\sim\tilde{t}^{2/[3(1+\tilde{w}_{\textrm{eff}})]}$
where the effective equation of state (EOS) is $\tilde{w}_{\textrm{eff}}=-1+\lambda^{2}/3=-1+\frac{2(2+n)^{2}}{9(1+n)^{2}}$.
This can be identified with the acceleration today if $\mu\sim H_{0}$.

Beside the final attractor, a coupled field with an exponential potential
has also another solution in which matter and field scale in the same
way with time and, consequently, their density fractions are constants.
This epoch has been denoted as $\phi$-matter-dominated era ($\phi$MDE)
\cite{Luca1}. As we will show in a moment, the $\phi$MDE plays a
central role in this work. This epoch occurs just after the radiation
era and replaces the usual MDE. During the $\phi$MDE the energy fraction
$\widetilde{\Omega}_{\phi}$ and the effective EOS $\widetilde{w}_{\textrm{eff}}$
are constant and given by \cite{Luca1,BNST} $\widetilde{\Omega}_{\phi}=\widetilde{w}_{\textrm{eff}}=4\beta^{2}/9\,.$
Then we have $\widetilde{\Omega}_{\phi}=\widetilde{w}_{\textrm{eff}}=1/9$
in $f(R)$ gravity theories, regardless of the form of $f(R)$. Therefore,
contrary to standard cosmology, in coupled models DE is not negligible
in the past (until the radiation era). In contrast to the accelerated
attractor, the $\phi$MDE occurs when the coupling term in the r.h.s.
of Eq.~(\ref{eq:phi}) dominates over the potential term, as it can
be explicitly shown. This aspect is crucial for the present work since
it implies that the $\phi$MDE exists independently of the form of
$f(R)$. In this regime the scale factor behaves as $\tilde{a}\sim\tilde{t}^{3/5}$:
in the JF this becomes $a\sim t^{1/2}$ instead of the usual $t^{2/3}$
behavior of the MDE. This is clearly a strong deviation from standard
cosmology and, as one can expect, is ruled out by observations, as
illustrated below. Notice that the JF evolution in this phase corresponds
to $R=0$ as during the radiation epoch but, just as in that case,
there is no singularity in an inverse power-law theory because this
behavior is not exact (see below). In the language of dynamical systems,
the $\phi$MDE is a saddle point.

To analyse numerically the full system (including the radiation energy
density $\rho_{\textrm{rad}}$ which obeys the standard conservation
equation in both frames), we introduce the following quantities: \begin{eqnarray}
x_{1}=\frac{\kappa\phi'}{\sqrt{6}},\quad x_{2}=\frac{\kappa}{\widetilde{H}}\sqrt{\frac{V}{3}},\quad x_{3}=\frac{\kappa}{\widetilde{H}}\sqrt{\frac{\rho_{\textrm{rad}}}{3}},\end{eqnarray}
 where a prime denotes the derivative with respect to $N\equiv{\textrm{log}}\,(\tilde{a})$.
The energy fractions of the field $\phi$ and of matter are given
by $\widetilde{\Omega}_{\phi}=x_{1}^{2}+x_{2}^{2}$ and $\widetilde{\Omega}_{m}=1-x_{1}^{2}-x_{2}^{2}-x_{3}^{2}$,
respectively. The effective EOS and the field EOS are $\widetilde{w}_{\textrm{eff}}=x_{1}^{2}-x_{2}^{2}+x_{3}^{2}/3$
and $\widetilde{w}_{\phi}=(x_{1}^{2}-x_{2}^{2})/(x_{1}^{2}+x_{2}^{2})$,
respectively. The complete system has been already studied in Ref.~\cite{Luca1}
and we will not repeat it here. The $\phi$MDE corresponds to the
fixed points $(x_{1},x_{2},x_{3})=(1/3,0,0)$ with $\widetilde{\Omega}_{\phi}=\widetilde{w}_{\textrm{eff}}=1/9$.
After this, the universe falls on the final attractor, which is the
accelerated fixed point $(x_{1},x_{2},x_{3})=(\lambda/\sqrt{6},\sqrt{1-\lambda^{2}/6},0)$
with $\widetilde{\Omega}_{\phi}=1$ and $\widetilde{w}_{\textrm{eff}}=-1+\lambda^{2}/3$.

To return to the JF we can simply apply the transformation law (\ref{relation}).
In the regime where radiation is negligible ($x_{3}\simeq0$) we obtain
the following effective EOS \begin{equation}
w_{\textrm{eff}}=\frac{1}{3}-\frac{2n}{3(1+n)}\frac{x_{2}^{2}}{(1-x_{1})^{2}}\,,\end{equation}
 together with the relation \begin{equation}
\Omega_{m}\equiv\frac{\kappa^{2}\rho_{m}}{3H^{2}F}=\frac{\widetilde{\Omega}_{m}}{(1-x_{1})^{2}}.\label{Omem}\end{equation}
 For the accelerated attractor we have $w_{\textrm{eff}}=-1+\frac{2(2+n)}{3(1+n)(1+2n)}$
(this relation was originally found in the context of inflation in
\cite{coa}), which gives $w_{\textrm{eff}}=-2/3$ for $n=1$. The
$\phi$MDE corresponds to $x_{2}=0$ and therefore $w_{\textrm{eff}}=1/3$
for any $n$, giving $a\propto t^{1/2}$. From Eq.~(\ref{Omem})
one has $\Omega_{m}=2$ and $\Omega_{\textrm{R}}\equiv(RF-f-6H\dot{F})/6FH^{2}=-1$
in the $\phi$MDE {[}see Eq.~(\ref{Jeq1}){]}. Since $\Omega_{\textrm{R}}$
does not need to be positive definite, $\Omega_{m}$ can be larger
than unity without any inconsistency.

Notice also that $w_{\phi}$ differs from the quantity $w_{DE}$ used
to annalyse SN data and defined through the equation $H^{2}=H_{0}^{2}[\Omega_{m}^{(0)}a^{-3}+(1-\Omega_{m}^{(0)})a^{-3}\exp(-3\int{\textrm{d}}a\, w_{\textrm{DE}}/a)]$.
We will also computte $w_{\textrm{DE}}$ below.

Most modifications of gravity suggested in the literature consider
terms in addition to the usual Einstein-Hilbert Lagrangian. For instance,
several authors have studied the following DE model \cite{Capo,Carroll}:
\begin{eqnarray}
f(R)=R-\mu^{2(n+1)}/R^{n}\,.\label{inverse}\end{eqnarray}
 In this case the potential in EF is given by \begin{eqnarray}
V(\phi)=Ae^{-\frac{2\sqrt{6}}{3}\kappa\phi}(e^{\frac{\sqrt{6}}{3}\kappa\phi}-1)^{\frac{n}{n+1}},\label{poEframe}\end{eqnarray}
 which vanishes at $\phi=0$ and has a maximum at $\kappa\phi=2(n+1)/(n+2)$.
In the limit $\phi\rightarrow\infty$ it behaves as $V(\phi)\propto\exp(-\lambda\kappa\phi)$.
For negative $n$ the pure exponential approximation is always good
during the past cosmic history if the higher-curvature term is responsible
for the present acceleration, because we are always in large $R$
limit. For positive $n$ the potential (\ref{poEframe}) differs from
the pure exponential one in the limit $R\gg\mu^{2}$ and one might
expect that for these values the evolution goes on as in the standard
case and, in particular, the $\phi$MDE disappears. However, we show
now by building an explicit solution that in reality this does not
happen.

Let us focus on the $R-\mu^{4}/R$ model taken for simplicity without
radiation. During the $\phi$MDE one can approximate the scale factor
in JF as $a(t)=(t/t_{i})^{1/2}+\epsilon(t)(t/t_{i})^{9/4}$ where
$t_{i}$ is an initial time at the beginning of the $\phi$MDE. This
solution is valid at first order in $\epsilon$ provided $\epsilon=(\mu^{2}/144H_{i}^{2})/((\rho_{m}^{(i)}/3H_{i}^{2})-\sqrt{H/H_{i}})^{1/2}$,
which is indeed small for $\mu$ of order $H_{0}$ as present acceleration
requires. After some time, the correction gets larger than the zero-th
order term itself and the $\phi$MDE is followed by a phase of accelerated
expansion. Then the beginning of the late-time acceleration is quantified
by the condition $t\approx(144H_{i}^{2}\sqrt{(\rho_{m}^{(i)}/3H_{i}^{2})}/\mu^{2})^{4/7}\, t_{i}$.
A similar argument applies for any $n<-1,n>-3/4$ with a correction
growing as $t^{5/2-1/2(n+1)}$. Since $R$ is of order $\mu^{2}$
at the beginning of the $\phi$MDE, the term $\mu^{4}/R$ dominates
over $R$ after the radiation era. This applies for any $\mu$, no
matter how small. In other words, the limit $\mu\to0$ of a fourth-order
theory does not reduce to second-order general relativity if, at the
same time, one imposes the conditions of acceleration today (cfr.
\cite{Fa06}).

For larger $\mu$ ($\gg H_{0}$) the $\phi$MDE can be shortened or
bypassed from the above condition, but then DE dominates soon without
a matter dominated epoch. Thus we have only two cases: either (i)
the $\phi$MDE exists, or (ii) a rapid transition from the radiation
era to the accelerating stage (without $\phi$MDE) takes place. In
summary, the system never behaves as in a standard cosmological scenario
except during radiation (during which matter and field play no role
in the expansion rate). In other words, whenever matter is dynamically
important, DE is also important as a consequence of the coupling.
We now confirm all this by a direct numerical integration.

In Fig.\,\ref{won1} we plot the evolution of $\Omega_{\phi}$, $\Omega_{\textrm{DM}}$
and $\Omega_{\textrm{rad}}$ and the equation of state in EF for $n=1,4$
and $10$ and $\mu\approx H_{0}$. The present values of the radiation
and matter density fractions are chosen to match the observations
in JF. As expected, the system enters the $\phi$MDE after the radiation
era and finally falls on the accelerated attractor. We ran our numerical
code for other positive and negative values of $n$ (from $n=-10$
to $n=10$, limiting to the accelerating cases) and found similar
cosmological evolutions. The plots in Fig.\,\ref{won1} are therefore
qualitatively valid for any $R+\alpha R^{-n}$ model (provided $F>0$).
It is also interesting to observe that the EOS is strongly varying
with time just near $z=0$: this should serve as a reminder that a
simple parametrization of $w_{{\rm DE}}$ may often fail to describe
interesting cosmologies.

We can show now that an effective EOS $w_{\textrm{eff}}=1/3$ during
the $\phi$MDE is cosmologically unacceptable. In principle this should
be shown case by case by a complete likelihood analysis of CMB and
LSS data (see \cite{aq} for such an analysis for various coupled
models), but this program is hardly feasible if we want to make general
statements on $f(R)$ theories. Instead we take a simpler but general
approach. We calculate the angular size of the sound horizon \begin{equation}
\theta_{s}=\int_{z_{\textrm{dec}}}^{\infty}\frac{c_{s}(z){\textrm{d}}z}{H(z)}\,/\int_{0}^{z_{\textrm{dec}}}\frac{{\textrm{d}}z}{H(z)}\,,\end{equation}
 where $c_{s}^{2}(z)=1/[3(1+3\rho_{b}/4\rho_{\gamma})]$ is the adiabatic
baryon-photon sound speed. 
According to the WMAP3y results \cite{WMAP}, the currently measured
value assuming a constant $w$ is $\theta_{s}=0.5946\pm0.0021$ deg.
As radiation follows the same conservation law, the thermal history
is the same as in usual cosmology so that $z_{\textrm{dec}}$ is unchanged.
It is easy to show that the integrand ${\textrm{d}}z/H(z)$ is conformally
invariant. For the model (\ref{inverse}) we integrate numerically
the equations of motion in EF (including radiation) by changing initial
conditions for $x_{1},x_{2},x_{3}$ via a trial and error procedure
until we obtain a present universe with a JF matter and radiation
densities as observed in the WMAP data (we used $\Omega_{m}^{(0)}=0.3$
and $h=0.7$). Once we have the full background solution we evaluate
$\theta_{s}$ with $z_{\textrm{dec}}$ obtained by solving the relation
$\widetilde{z}_{\textrm{dec}}e^{\kappa(\phi-\phi_{0})/\sqrt{6}}=z_{\textrm{dec}}$.
We have two competing effects: first, since the $\phi$MDE is more
decelerated than in the usual case $r(z_{\textrm{dec}})$ will be
systematically smaller; second, the physical sound horizon distance
at decoupling is smaller than in usual cosmology partly because in
our models $a\propto t^{1/2}$ also between $t_{\textrm{eq}}$ and
$t_{\textrm{dec}}$ with $z_{\textrm{eq}}>z_{\textrm{dec}}$ and mostly
because $H(z)$ before decoupling is much higher than in the standard
case. Assuming $H(z)=H_{0}(1+z)^{2}$ instead of $(1+z)^{3/2}$, $H(z_{dec})$
becomes 30 times larger than in standard models. We find that the
second effect is by far the dominating one, and as a consequence $\theta_{s}$
turns out to be an order of magnitude smaller than the observed value.
In practice, we find that $\theta_{s}$ can be approximated to a few
percent using a standard cosmological model with an uncoupled dark
energy component and a matter component with effective equation of
state $w=1/3$. The typical values we find are near $\theta_{s}\approx0.03$
deg, i.e. more than ten times smaller than in a standard model. The
periodic spacing $\Delta\ell$ between the acoustic peaks in the CMB
will be larger too by nearly the same factor.

\begin{figure}
\includegraphics[%
  scale=0.35]{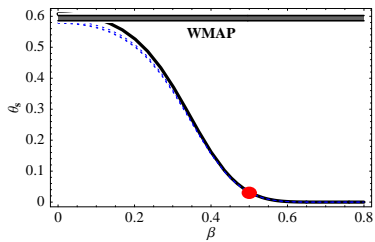}

\caption{The sound horizon angular distance $\theta_{s}$ as a function of
the coupling $\beta$ for $n=1$ (thick line) and $n=-2,3,10$ (dotted
lines). The disk marks the value for $\beta=1/2$. The grayed region
shows the WMAP3y constraint at $4\sigma$.\label{distance}}
\end{figure}

In Fig. \ref{distance} we plot the value of $\theta_{s}$ as a function
of the coupling constant $\beta$ (see eqs. \ref{eq:phi}-\ref{eq:rhom})
and for several $n$'s. Clearly there is no way that small changes
in $\Omega_{m}^{(0)}$ or $h$ or $w_{\mathrm{eff}}$ can cure this
problem. Discrepancies are found as well for the age of the universe
which turns out to be near 10-11 Gyr. As to be expected, we also find
that the perturbations depart significantly from the standard case
(as in pure exponential coupled models \cite{aq}): for the matter
density contrast on sub-horizon scales we find $\delta\sim a^{2}$
during $\phi$MDE instead of the standard linear law.

It is possible to generalize our results in several ways. First, one
can show by direct substitution that the standard matter era $t^{2/3}$
is a solution of (\ref{Jeq1}) only for pure power laws (plus possibly
a cosmological constant) $R^{-n}$ with $n=-1,-(7\pm\sqrt{73})/12$.
In the last two cases, however, the {}``matter'' era occurs for
$\Omega_{m}=0$ and is therefore unacceptable. The first case is clearly
the pure Einstein case: this shows that a standard sequence of (exact)
$t^{2/3}$ expansion followed by acceleration can occur only for $\Lambda$CDM.
In contrast, the $\phi$MDE generically exists as a saddle point.
For $R^{-n}$ with $-1<n<-3/4$ the $\phi$MDE is instead a stable
point and the models are ruled out anyway. Still, this alone does
not guarantee that the $\phi$MDE is always reached, regardless of
the initial conditions, and a case-by-case numerical analysis is necessary.
We explored extensively models like $f(R)=\exp(R)$ and $\log(R)$
and always found $\phi$MDE before acceleration. We also carried out
a preliminary analysis of Lagrangians for the models like $f=R-c(R_{\mu\nu}R^{\mu\nu})^{-n}$,
$f=R-c(R_{\alpha\beta\gamma\delta}R^{\alpha\beta\gamma\delta})^{-n}$
and $f=R-c(R_{{\rm GB}})^{-n}$ (where $R_{{\rm GB}}$ is a Gauss-Bonnet
term) and did not find any acceptable cosmological evolution.

For the $f(R)=R-\mu_{1}^{4}/R+\mu_{2}R^{2}$ models, two mechanisms
that could satisfy the local gravity constraints were suggested. One
\cite{ON} achieves a short interaction range (or a large field mass)
by adjusting $\mu_{1},\mu_{2}$ so that ${\rm d}^{2}V(\phi)/{\rm d}\phi^{2}$
vanishes today, when $R=\sqrt{3}\mu_{1}^{2}$. Before this the $R^{2}$
term dominates and therefore we are back in one of our cases and the
$\phi$MDE takes place. Moreover, we find that the local minimum in
the EF potential for such a class of Lagrangians does not lead to
a late-time (effective) cosmological constant. Another possibility
to build a large mass is to take a very small $\mu_{2}$ \cite{brook},
but in this case it is the $1/R$ term that dominates the cosmological
evolution from the end of radiation, and again we are back in one
of our cases. So even models that are designed to pass local gravity
experiments fail our cosmological test. In summary, the main feature
of our analysis is the modification of the standard matter dominated
epoch for the $f(R)$ dark energy models investigated here. Hence
these models are ruled out as viable cosmologies even if they are
arranged to pass the Supernovae test and the local gravity constraints.
We conjecture that our results apply to a much larger class of $f(R)$
models; the precise conditions that determine the cosmological behavior
will be published in future works. 

\vspace{0.5cm}
\textit{Acknowledgments}-- L.\,A. acknowledges the hospitality at
the Gunma College of Technology and support from JSPS. The work of
S.\,T. is supported by JSPS.

\end{document}